\documentclass[twoside,titlepage]{article}
\usepackage{color}

\usepackage{amsmath, amsthm, amssymb} 
\usepackage{mathrsfs}
\usepackage{amsfonts}
\usepackage{hyperref} 
\usepackage{graphicx} 
\numberwithin{equation}{section} 

\DeclareMathAlphabet{\mathpzc}{OT1}{pzc}{m}{it}
\DeclareMathAlphabet{\mathpzc}{OT1}{pzc}{m}{it}

\begin{document}


\newcommand{\be}{\begin{equation}}
\newcommand{\e}{\end{equation}}
\newcommand{\bea}{\begin{equation*}}
\newcommand{\ea}{\end{equation*}}
\newcommand{\la}{\label}
\newcommand{\bu}{\bullet}

\begin{titlepage}

\today

\begin{center}

\hfill{\tt WIS/02/13-APR-DPPA}\\


\vskip 20mm

{\Large{\bf Local gravitational solutions dual to M2-branes intersecting and/or ending on M5-branes}}

\vskip 10mm

{\bf Leon~Berdichevsky and Bat-el~Dahan\footnote{Emails: leon.berdichevsky, batel.dahan@weizmann.ac.il}}

\vskip 4mm
{\em Department of Particle Physics and Astrophysics,}\\
{\em Weizmann Institute of Science,}\\
{\em Rehovot 76100, Israel}\\
[2mm]

\end{center}
\vskip 2cm

\begin{center} {\bf ABSTRACT }\end{center}
\begin{quotation}
\noindent

We construct local solutions to eleven-dimensional supergravity describing M2-branes intersecting and/or ending on M5-branes, in the near-horizon limit of the M2-branes. Global solutions describing these theories should have 16 supercharges and asymptotic $AdS_4 \times S^7$ regions, as well as M5-branes wrapping $AdS_3 \times S^3$ subspaces. We construct the local solution describing a stack of an arbitrary number of M5-branes with $AdS_3 \times S^3$ worldvolume and arbitrary M2-brane charge. Our construction provides a mechanism to get around the no-go theorem 
that rules out the existence of global solutions preserving 16 supercharges interpolating between $AdS_7 \times S^4$ and $AdS_4 \times S^7$ regions.

\end{quotation}

\vfill

\end{titlepage}

\tableofcontents

\section{Introduction and summary of results}

The AdS/CFT correspondence \cite{Maldacena:1997re,Witten:1998qj,Gubser:1998bc}
has provided a powerful theoretical framework to study conformal field theories (CFTs) using classical gravitational physics. Examples of the original correspondence \cite{Maldacena:1997re}
 are the duality between type IIB string theory on $AdS_5 \times S^5$ and 4d $U(N)$ $\mathcal N=4$ SYM, and the duality between M-theory on $AdS_4 \times S^7$ ($AdS_7 \times S^4$) and the low-energy theory on a stack of M2(M5)-branes. The former arises as the near-horizon (or decoupling) limit of a stack of D3-branes, while the latter is realized as the decoupling limit of a stack of M2(M5)-branes. The low-energy theory on the M2-branes is the recently formulated ABJM theory with $k=1$ \cite{Aharony:2008ug}, while the low-energy theory on the M5-branes is a 6d (2,0) supersymmetric theory that is still not fully understood. A systematic approach to construct new dual pairs of theories is to consider more complicated brane configurations and take the decoupling limit for one set of branes, since many interesting quantum field theories arise as the low-energy limit of branes intersecting  and ending on other branes (following \cite{Ganor,Hanany:1996ie}).

Finding the gravitational solutions describing configurations of branes intersecting other branes,
or branes ending on other branes, is a challenging problem, which only has a solution in some very special cases. Recently, there has been progress in the construction of solutions to type IIA \cite{Lin:2004nb,Gaiotto:2009gz,ReidEdwards:2010qs,Aharony:2012tz} and to type IIB supergravity \cite{D'Hoker:2007xy,D'Hoker:2007xz,Aharony:2011yc,Assel:2011xz,Benichou:2011aa,Assel:2012cj} invariant under 16 supercharges that describe configurations of branes intersecting and/or ending on other branes, in the near-horizon limit of one set of branes; all the remaining branes remain localized. These solutions are relevant for the AdS/CFT correspondence, and therefore it is natural to extend these results by searching for gravity solutions dual to configurations of branes intersecting and/or ending on other branes in M-theory, that preserve the same amount of supersymmetry. In this paper we study this problem for M2-branes intersecting and/or ending on M5-branes, in the near-horizon limit of the M2-branes.

The intersection with the M5-branes breaks half of the 32 supersymmetries of the low-energy theory on the M2-branes, and breaks its bosonic
$SO(3,2) \times SO(8)$ symmetry to $SO(2,2) \times SO(4)\times SO(4)$. An important step for the construction of solutions dual to M2-branes intersecting and/or ending on M5-branes was taken in \cite{Yamaguchi:2006te,Lunin:2007ab,D'Hoker:2008wc,Estes:2012vm}, where the general local solution to 11d supergravity with the isometry $SO(2,2) \times SO(4)\times SO(4)$ and invariant under 16 supercharges was constructed. The solution is a warped $AdS_3 \times S^3 \times S^3$ space over a Riemann surface $\Sigma$ with boundary, and has non-trivial 3-form potential. There is a one parameter family of supergroups with 16 supercharges and with $SO(2,2) \times SO(4)\times SO(4)$ bosonic subgroup \cite{Estes:2012vm}. Solutions locally asymptotic to $AdS_7 \times S^4$ are invariant under $OSp(4^*|2) \otimes OSp(4^*|2)$. Globally regular solutions with one asymptotic $AdS_7 \times S^4$ region were constructed in \cite{D'Hoker:2008qm}. These solutions are dual to the supersymmetric self-dual string soliton solution of the 6d (2, 0) supersymmetric M5-brane worldvolume theory. On the other hand, solutions locally asymptotic to $AdS_4 \times S^7$ preserve the supergroup $OSp(4|2,\mathbb{R}) \otimes OSp(4|2,\mathbb{R})$. A class of physically acceptable global solutions with two asymptotic $AdS_4 \times S^7$ regions was obtained in \cite{D'Hoker:2009gg}, and referred to as the M-Janus solution. It consists of a one-parameter family of deformations of $AdS_4 \times S^7$ which gives a holographic realization of a Janus-like defect/interface M2-brane theory. The absence of deformations of $AdS_4 \times S^7$ beyond those of the M-Janus solution was then proven in \cite{D'Hoker:2009my}. It follows that the symmetries of these theories imply a no-go theorem: solutions preserving 16 supercharges with both $AdS_4 \times S^7$ and $AdS_7 \times S^4$ asymptotic regions don't exist\footnote{This two-branch structure for 1/2-BPS solutions with $SO(2,2) \times SO(4)\times SO(4)$ isometry was first discovered in \cite{Lunin:2007ab}. A similar result showing the existence of different branches of 1/2-BPS solutions with the isometry $SO(4,2) \times SO(3)$ was found in \cite{OColgain:2012wv}.}, which seems to be an obstruction for the construction of solutions dual to M2-branes intersecting and/or ending on M5-branes.

The analogy with other cases of branes intersecting and/or ending on other branes in type IIA \cite{Lin:2004nb,Gaiotto:2009gz,ReidEdwards:2010qs,Aharony:2012tz} and type IIB string theory \cite{D'Hoker:2007xy,D'Hoker:2007xz,Aharony:2011yc,Assel:2011xz,Benichou:2011aa,Assel:2012cj} suggests that the 11d supergravity solutions dual to M2-branes intersecting and/or ending on M5-branes should include singularities arising from M5-branes sitting
in the background. The simplest global solutions should have two $AdS_4 \times S^7$ regions and one M5-brane singularity for M2-branes intersecting and ending on a single stack of M5-branes, or one $AdS_4 \times S^7$ region with one M5-brane singularity for M2-branes ending on a single stack of M5-branes. General global solutions should have several stacks of M5-branes wrapping different 3-spheres, with the stacks localized at different points on the boundary of the Riemann surface $\Sigma$. For solutions with only one $AdS_4 \times S^7$ region, there should be a one-to-one mapping between global solutions and supersymmetric boundary conditions for M2-branes ending on M5-branes \cite{Berman:2009kj,Chu:2009ms,Berman:2009xd}, analogous to the mapping between the global solutions of type IIB supergravity found in \cite{Aharony:2011yc} and boundary conditions for D3-branes ending on 5-branes \cite{Gaiotto:2009gz}.

In the present paper, we find the local description of the M5-brane singularity. Specifically, we construct a local solution describing a localized stack of an arbitrary number of M5-branes and invariant under the $OSp(4|2,\mathbb{R}) \otimes OSp(4|2,\mathbb{R})$ supergroup. The M5-branes wrap an $AdS_3 \times S^3$ subspace and possibly carry M2-brane charge. Our construction provides a mechanism for the existence of solutions describing M2-branes intersecting and/or ending on M5-branes, in the near-horizon limit of the M2-branes, consistent with the no-go theorem described above. Uplifting these local solutions to global ones is still an open problem.

The rest of the paper is organized as follows: In section 2 we review local solutions to 11d supergravity invariant under the $OSp(4|2,\mathbb{R}) \otimes OSp(4|2,\mathbb{R})$ supergroup. We then review two global solutions - the $AdS_4 \times S^7$ solution itself, and the M-Janus solution which has two asymptotic $AdS_4 \times S^7$ regions. In section 3 we show that a local solution can have only one more type of singularity besides the singularity describing an $AdS_4 \times S^7$ asymptotic space. In section 4 we show that this other type of singularity describes a stack of an arbitrary number of M5-branes wrapping an $AdS_3 \times S^3$ subspace and possibly carrying M2-brane charge. We interpret these local solutions as describing M2-branes intersecting and/or ending on M5-branes, in the near-horizon limit of the M2-branes, and in the region close to the M5-branes.

\section{Review of M-theory solutions with $OSp(4|2,\mathbb{R}) \otimes OSp(4|2,\mathbb{R})$ symmetry}

\subsection{11d supergravity with $SO(2,2) \times SO(4)\times SO(4)$ isometry}

The general solution to 11d supergravity with the isometry $SO(2,2) \times SO(4)\times SO(4)$ that preserves 16 supercharges was found in \cite{D'Hoker:2008wc,Estes:2012vm}. The bosonic symmetry requires a space-time with the geometry $AdS_3\times S^{3}_2\times S^{3}_3\times \Sigma$ warped over a Riemann surface $\Sigma$ with boundary $\partial\Sigma$. The metric takes the form
\be
ds^2 = f_1^2 ds^2_{AdS_3} + f_2^2 ds^2_{S^3_2} + f_3^2 ds^2_{S^3_3} + \rho^2 ds^2_{\Sigma},
\e
where $f_1, f_2, f_3$ and $\rho$ are real functions on $\Sigma$.
The solution has non vanishing 3-form potential
\be
C_3 = b_1 \hat \omega_{AdS_3} +b_{2} \hat \omega_{S^3_2} +b_{3} \hat \omega_{S_3^3},
\e
with the associated conserved 4-form field strength $F_4 = dC_3$ given by
\be
\begin{split}
F_4= & g_{1a}\omega_{AdS_3} \wedge e^{a}+g_{2a} \omega_{S^3_2} \wedge e^{a}+g_{3a} \omega_{S_3^3} \wedge e^{a}. \\
\end{split}
\e
We use the notation $\hat \omega_{AdS_3}$ ($\omega_{AdS_3}$) and $\hat \omega_{S^3_{2,3}}$ ($\omega_{S^3_{2,3}}$) for the volume forms on the unit (warped) $AdS_3$ and $S^3_{2,3}$ respectively, and $e^a, a=1,2$ is an orthonormal frame on $\Sigma$. The real functions $b_i$ and $g_{ia}$ take values on $\Sigma$ and satisfy $ \partial_a b_i = - f_i^3 g_{ia}$.

\subsection{Local solutions to the BPS equations}

In \cite{D'Hoker:2008wc} the BPS equations were reduced to a simple partial differential equation (PDE)
\be \label{eq:PDE}
\partial_{w} G=\dfrac{1}{2}(G+\bar G)\partial_{w}\ln(h),
\e
where $h$ is a real, positive harmonic function, $G$ is a complex function and $(w, \bar w)$ are arbitrary complex coordinates on $\Sigma$.

It is sometimes convenient to write $h$ in terms of an holomorphic function $H(w)$ as follows\footnote{The analysis in \cite{D'Hoker:2009my} makes use of a slightly different holomorphic function $\kappa = 2i H$.}
\be
h= H(w) + \bar H(\bar w),
\e
and to express $G$ in terms of a real function $\Phi$ and $H$
\be
G=\dfrac{ \partial_{\bar w} \Phi}{\partial_{\bar w} \bar H}.
\e
The expressions simplify if we choose the local coordinate system $w=H(w)$. In terms of $H$ and $\Phi$, the PDE becomes
\be \label{eq:PDE.Phi}
2(H + \bar H)\partial_{H}\partial_{\bar H}\Phi-\partial_{\bar H}\Phi-\partial_H\Phi=0.
\e

The supergravity fields are expressed in terms of $h$ and $G$ (or equivalently $H$ and $\Phi$) and depend on the supergroup preserved by the solution. We will focus on solutions invariant under $OSp(4|2,\mathbb{R})\otimes OSp(4|2,\mathbb{R})$, as this is the supergroup preserved by solutions locally asymptotic  to $AdS_4 \times S^7$ \cite{D'Hoker:2009gg,D'Hoker:2009my}. In order to write the supergravity fields in a compact form we use the notation $W^2=4|G|^4+(G-\bar G)^2$. The metric factors are expressed as follows
\be
\begin{array}{ll}
f_1^6= \dfrac{h^2W^2}{16^2(|G|^2-1)^2}, &  \quad  f_2^6= h^2\dfrac{(|G|^2-1)}{4W^4}(2|G|^2+i(G-\bar G))^3,\\
& \\
f_3^6= h^2\dfrac{(|G|^2-1)}{4W^4}(2|G|^2-i(G-\bar G))^3, & \quad  \rho^6= \dfrac{|\partial h|^6}{16^2h^4}(|G|^2-1)W^2.\\
\end{array}
\e
Similarly, the functions defining the 4-form field strength are given by
\be
\begin{split}
-\partial_w b_1 = f^3_1g_{1w} = & \dfrac{3W^2\partial_wh}{32G(|G|^2-1)}-\dfrac{1+G^2}{16G(|G|^2-1)^2}J_w,\\
-\partial_w b_2= f^3_2g_{2w} = & -\dfrac{(G+i)(2|G|^2+i(G-\bar{G}))^2}{W^4}J_w,\\
-\partial_w b_3= f^3_3g_{3w} = & \dfrac{(G-i)(2|G|^2-i(G-\bar{G}))^2}{W^4}J_w,\\
\end{split}
\e
where we defined the vector
\be
J_w=\dfrac{1}{2}(G\bar G-3\bar G^2+4G\bar G^3)\partial_w h+hG\partial_w \bar G.
\e
The functions $h$ and $G$ are constrained by local regularity and boundary conditions. Local regularity imposes the constraint
\be \label{eq:reg}
|G|^2>1 \quad  \textrm{for all $(w, \bar w)$ in the interior of $\Sigma$}.
\e
On the other hand, the Riemann surface $\Sigma$ has boundaries but the full 11d space
does not, and thus some cycle always has to shrink at the boundary $\partial \Sigma$. Therefore, one of the metric factors for the 3-spheres should vanish at $\partial \Sigma$ while the rest of the metric factors should remain finite. This condition translate into the constraints
\be \label{eq:b.c.}
h=0, \text{ and $G=+i$ or $G=-i$ for all $(w,\bar w)$ in $\partial\Sigma$},
\e
except for isolated points. Note that the boundary conditions \eqref{eq:b.c.} constrain $\textrm{Re}(H) \ge 0$.

\subsection{Global solutions}

\subsubsection{The  $AdS_4\times S^7$ solution}

The simplest solution to  \eqref{eq:PDE}, subject to the regularity condition \eqref{eq:reg} and the boundary conditions \eqref{eq:b.c.}, is the maximally symmetric solution $AdS_4\times S^7$ \cite{D'Hoker:2008wc}. The Riemann surface $\Sigma$ has the topology of the disk and is conveniently parameterized by an infinite strip $\Sigma=\{w\in \mathbb{C}|w=x+iy,x\in\mathbb{R},y\in[0,\pi/2]\}$. In terms of these coordinates, the functions $h$ and $G$ take the following form
\begin {equation}
h=8 \eta \text{Im}\left(\sinh(2w)\right),  \qquad	 G=i\dfrac{\cosh(w+\bar w)}{\cosh(2\bar w)},
\end{equation}
where $\eta$ is a positive constant.

The boundary is characterized by the vanishing of the harmonic function $h = 0$, and $G = \pm i$. On the lower boundary of the strip, where $y = 0$, one has $ G = i$,
which implies that the radius $f_2$ of $S_2^3$ vanishes. On the upper boundary of the strip, where $y =\pi/2$, one has $G = -i$, which implies that the radius $f_3$ of $S^3_3$ vanishes.
The corresponding metric is
\begin {equation}
ds^2=\eta^{2/3} \cosh^2(2x)ds^2_{AdS_3}+4 \eta^{2/3} dx^2+4 \eta^{2/3}\left(\cos^2(y)ds^2_{S^3_2}+\sin^2(y)ds^2_{S^3_3}+dy^2\right).
\end{equation}
We can see how the $AdS_4\times S^7$ structure arises - the $AdS_3$ combines with the $x$-coordinate to form an $AdS_4$, while the two 3-spheres and the compact $y$ direction form the $S^7$. The radii are given by $R_{S^7} = 2 R_{AdS_4}= 2 \eta^{1/3}$.

The 4-form field-strength components are
\be
g_{1w} = -3, \qquad g_{2w} = 0, \qquad g_{3w} = 0.
\e
The non-zero 7-form flux supported by the $S^7$ is
\be
\int_{S^7} \ast F_{4} = 2^3 \pi^4 \eta^2 \equiv (2 \pi)^6 N_2.
\e
We therefore recover the near-horizon limit of $N_2$ M2-branes.

\subsubsection{The M-Janus solution} \label{subsec:AdS4xS7}

A special class of regular solutions is given by the M-Janus solution \cite{D'Hoker:2009gg}. The M-Janus solution is a one-parameter deformation of the $AdS_4 \times S^7$ solution. Thus, the Riemann surface $\Sigma$ has the topology of the disk. For future analysis, we choose to characterize the solution by the holomorphic function $H$ and the real function $\Phi$ on $\Sigma$, and to parameterize the disk by the upper half-plane $\Sigma = \{ u \in \mathbb{C}| \textrm{Im}(u) \ge 0 \}$. The map between the upper half-plane  and the infinite strip considered above is given by $u = \tanh(w)$. In this coordinate system the functions take the form
\be
\begin{split}
H(u)= & 4 i \eta \left( \frac{1}{u+1} + \frac{1}{u-1} \right), \\
\Phi(u, \bar u) = & - (\lambda + i) \left[H(u) + \frac{1}{2}\right]^{1/2}\left[\bar H(\bar u) - \frac{1}{2}\right]^{1/2} + c.c.\\
= & 8 \eta \frac{u + \bar u - \lambda (u \bar u -1)}{|u^2 -1|}
\end{split}
\e
where $\eta$ is a positive constant and $\lambda$ is the deformation parameter, which can take any real value. The maximally symmetric solution is recovered when $\lambda=0$.
Note that there are two special points $ u = \pm 1$ at which $H = \infty$. These points are located at the boundary of $\Sigma$. We show next that near each of these two points the solution asymptotes to an $AdS_4 \times S^7$ region.

\subsubsection*{Near the $AdS_4 \times S^7$ regions}

We now show that at the point $ u =  1$ the geometry is locally asymptotic to $AdS_4 \times S^7$. To do this we consider polar coordinates $ u = r e^{i \theta} -1$ and take the limit $ r \rightarrow 0$. In this limit, the functions $h$ and $\Phi$ take the form
\be
\begin{split}
\frac{1}{8 \eta} h \rightarrow & \frac{1}{r} \sin(\theta) + \frac{1}{4}r \sin(\theta)+\frac{1}{8}r^2 \sin(2 \theta)+\frac{1}{16}r^3 \sin(3\theta),\\
\frac{1}{8 \eta} \Phi \rightarrow & - \frac{1}{r} + \left(\frac{1}{2}+ \lambda \right) \cos(\theta)+ \left[ \left(\frac{3}{16}-\frac{1}{4}\lambda\right) + \left(\frac{1}{16}+\frac{1}{4} \lambda \right) \cos (2 \theta) \right] r \\
& + \left[ \left(\frac{3}{32}-\frac{3}{16}\lambda\right) + \left(\frac{1}{32}+\frac{3}{16} \lambda \right) \cos (2 \theta) \right] r^2 \cos(\theta) \\
& + \left[ \left(\frac{15}{1024}-\frac{1}{128}\lambda\right) + \left(\frac{11}{256}-\frac{1}{32} \lambda \right) \cos (2 \theta) + \left(\frac{5}{1024}+\frac{5}{128} \lambda \right) \cos (4 \theta) \right] r^3,\\
\end{split}
\e
The associated function $G$ behaves as
\be
\begin{split}
G \rightarrow & e^{-i \theta} \left\{-i - \left(\frac{1}{2} + \lambda\right)r \sin(\theta)+ \left[ \left(\frac{1}{4}-\lambda\right) \cos(\theta) + i \left(- \frac{3}{8}+\frac{1}{2} \lambda \right) \sin (\theta) \right] r^2 \sin(\theta) \right. \\
& \left.  + \left[ \left(\frac{3}{32}-\frac{1}{16}\lambda\right) + \left(\frac{17}{32}-\frac{5}{16} \lambda \right) \cos (2 \theta) + 2i \left(-\frac{9}{32}+\frac{1}{16} \lambda \right) \sin (2 \theta) \right] r^3 \sin(\theta) \right\},\\
\end{split}
\e
The metric factors then have the following asymptotic behavior
\be
\begin{array}{ll}
f_1^2 \rightarrow \eta^{2/3} \frac{1}{(1+ \lambda^2)^{2/3}} \frac{1}{r^2}, &  \quad  f_2^2 \rightarrow 4 \eta^{2/3} (1+ \lambda^2)^{1/3} \cos^2\left(\frac{\theta}{2}\right),\\
& \\
f_3^2 \rightarrow 4 \eta^{2/3} (1+ \lambda^2)^{1/3} \sin^2\left(\frac{\theta}{2}\right), & \quad  \rho^2 \rightarrow \frac{1}{4} \eta^{2/3} (1+ \lambda^2)^{1/3} \frac{1}{r^2}.
\end{array}
\e
Upon the change of coordinates
\be
r \rightarrow \frac{r}{(1+ \lambda^2)^{1/2}}, \qquad \theta \rightarrow 2\theta,
\e
one obtains the canonical $AdS_4 \times S^7$ metric with radius $R_{S^7} = 2 R_{AdS_4}=2 \eta^{1/3}(1+\lambda^2)^{1/6}$.

To compute the 4-form field strength we need the vector $J_{u}$. It takes the form
\be
J_u \rightarrow - 16 \eta \lambda (1+\lambda^2) r \sin^3(\theta).
\e
Note that the first three orders in the expansion cancel out in $J_u$. The components of the 4-form field strength become
\be
g_{1u} \rightarrow \frac{3}{2}\frac{1}{r}e^{-i\theta}, \qquad g_{2u} \rightarrow -2 \lambda (1 +\lambda^2)^{1/2} r e^{-i\frac{\theta}{2}}, \qquad g_{3u} \rightarrow i 2 \lambda (1 +\lambda^2)^{1/2} r e^{-i\frac{\theta}{2}}.
\e
The solution carries non-zero 7-form flux through $S^7 \simeq \left(S^3_2 \times S_3^3\right) \times_f [0,\pi]_{\theta}$
\be
\int_{S^7} \left( \ast F_4 + \frac{1}{2} C_3 \wedge F_4 \right) = 2^3 \pi^4 \eta^2 (1+\lambda^2) \equiv (2 \pi)^6 N_2.
\e
The solution is therefore locally asymptotic to the near-horizon geometry of a stack of $N_2$ M2-branes.

A similar result holds near the point $u=-1$ with the 3-spheres $S_2^3$ and $S_3^3$ interchanged.


\section{Allowed local singularities}

We reviewed in the last section a one-parameter family of regular solutions, the M-Janus solution, and showed that it contains two singularities of the same type in the metric. Near each singularity there is  an $AdS_4 \times S^7$ region, the near-horizon geometry of a stack of M2-branes. These singularities are necessarily located at the points at which the function $H$ blows up. We expect to find another type of singularity coming from the near-horizon limit of M2-branes intersecting and/or ending on M5-branes, which should locally look like an M5-brane, in analogy to what was found for the near-horizon limit of D3-branes intersecting and/or ending on 5-branes \cite{D'Hoker:2007xy,D'Hoker:2007xz,Aharony:2011yc}. We show in this section that this singularity is indeed allowed, and in fact it is the only other allowed singularity.

We are interested in the region where $H\rightarrow\infty$. We analyze the local behavior near this point using polar coordinates
$\{0 \le r < \infty,\theta\in[0,\pi]\}$ such that $H=\infty$ is mapped to $r=0$. Now, the function $h$ is real, non-negative and harmonic. Reality and harmonicity imply that it can be written in the most general form as the following series
\begin{equation}
h= \sum_{n=-\infty}^{\infty} \left[ a_n \dfrac{\sin(n\theta)}{r^n}+\tilde{a}_n \dfrac{\cos(n\theta)}{r^n} \right],
\end{equation}
where $a_n, \tilde{a}_n$ are arbitrary real constants.
In the given range of $\theta$, only $\sin(\pm \theta)$ is non-negative everywhere. Thus, positivity of $h$ in the interior of $\Sigma$ dictates the following form near $r=0$
\begin {equation}
\begin{aligned}
h&= a_{-1} \dfrac{\sin(\theta)}{r} + a_1 r \sin(\theta) + O(r^2),\qquad H =  \frac{i}{2} a_{-1} \dfrac{e^{-i\theta}}{r}+ \frac{i}{2}  c - \frac{i}{2}  a_1 r e^{i \theta}+ O(r^2),
\end{aligned}
\end{equation}
where $a_{-1}>0$, $c$ is a real constant and we used the boundary condition \eqref{eq:b.c.} on $h$ to fix $\tilde a_n=0$ for all $n$.

The boundary condition \eqref{eq:b.c.} on $G$ requires $H$ and $\Phi$ to have the same degree of singularity near $r=0$. So, the most general form of $\Phi$ near $r=0$ is
\begin {equation}
 \Phi=\dfrac{f(\theta)}{r}+g(\theta)+m(\theta)r+ O(r^2).
\end{equation}
Substituting the forms of $H$ and $\Phi$ into the differential equation \eqref{eq:PDE.Phi}, we get
\begin{equation} \label{eq:coeff}
\begin{aligned}
f(\theta)&=\mathcal A\cos(\theta)+\mathcal B, \qquad g(\theta)= A \cos(\theta)+ B, \\
m(\theta)&= \left(\frac{1}{2} C -\frac{a_1}{a_{-1}} \mathcal B \right) \cos(2\theta) + \left(D - \frac{a_1}{2 a_{-1}}\mathcal A \right) \cos(\theta) - \frac{1}{2} C + \left(D + \frac{a_1}{2 a_{-1}}\mathcal A \right) \times \text{log terms},
\end{aligned}
\end{equation}
where $\mathcal A, \mathcal B, A, B, C, D$ are integration constants.

In order to satisfy the boundary condition \eqref{eq:b.c.} on $G$, we have only two choices: $\mathcal A=0$ or $\mathcal B=0$. We must also set the coefficient of the logarithmic terms to zero $2 a_{-1} D = - a_1 \mathcal A$. Without loss of generality we choose $B=0$ since the physical quantity $G$ doesn't contain this term.

The first choice $\mathcal A=0$ corresponds to an asymptotic $AdS_4 \times S^7$ region as shown in detail in sec \ref{subsec:AdS4xS7}. In this case the boundary conditions \eqref{eq:b.c.} fix $\mathcal B = -a_{-1}$, and the result follows upon the identification
\be
a_{-1} = 8 \eta, \qquad a_1 = 2 \eta, \qquad A=4 \eta \left(1+2\lambda \right), \qquad C=-\eta(3-4\lambda).
\e

In the next section we show that the second choice $\mathcal B=0$ corresponds to a local solution describing a stack of M5-branes wrapping an $AdS_3 \times S^3$ space and possibly carrying M2-brane charge.

\section{M2-branes intersecting and/or ending on M5-branes: Local solutions}

We show in this section that there are local solutions describing a stack of M5-branes wrapping an $AdS_3 \times S^3_i$ subspace, with either $i=1,2$, and possibly carrying M2-brane charge. The stack is located at $H=\infty$ and the local solution is given by the second choice $\mathcal B=0$ in \eqref{eq:coeff}. If we parameterize $\Sigma$ near this point using polar coordinates $\{0 \le r < \infty,\theta\in[0,\pi]\}$ such that $H=\infty$ is mapped to $r=0$, the local solution as $r \rightarrow 0$ reads as
\be
\begin{split}
h  \rightarrow & a \frac{1}{r} \sin(\theta)+ a_1 r \sin(\theta), \qquad H \rightarrow \frac{i}{2} \left[a  \frac{1}{r} e^{- i \theta} + c - a_1 r e^{i \theta} \right],\\
\Phi \rightarrow & a \frac{1}{r} \cos(\theta) + A \cos(\theta) + C r \sin^2(\theta) - a_1 r \cos(\theta),\\
\end{split}
\e
where we relabeled $ a_{-1} \equiv a >0$ and used the boundary conditions \eqref{eq:b.c.} to fix $\mathcal A =a$. The real constants $A$, $C$ and $c$ are arbitrary. The associated function $G$ behaves as
\be \label{eq:Gs}
\begin{split}
G & \rightarrow i - \frac{A}{a} \sin(\theta) e^{-i \theta} r + \frac{C}{a} \left[e^{-i \theta} \cos(\theta) +  e^{-i 2 \theta}\right] \sin(\theta) r^2\\
\end{split}
\e
and satisfies $G=i$ at the boundary. The corresponding expression for $G$ satisfying the boundary condition $G=-i$ is given by the complex conjugate of \eqref{eq:Gs}. The two boundary conditions correspond to different cycles vanishing on the boundary of $\Sigma$; $S^3_2$ for $G=i$, and $S^3_3$ for $G=-i$.
The regularity condition $|G|>1$ requires $A>0$.

The metric factors take the form
\be
f_1^6 \rightarrow \frac{1}{2^7}\frac{a^3}{A} \frac{1}{r^3} , \qquad  f_2^6 \rightarrow \frac{A^2}{2^4}  \sin^6(\theta),  \qquad f_3^6 \rightarrow \frac{1}{2}\frac{a^3}{A} \frac{1}{r^3}, \qquad \rho^6 \rightarrow \frac{A^2}{2^{10}} \frac{1}{r^6}.
\e
such that
\be \label{eq:M5.metric}
ds^2 \rightarrow   \frac{1}{2^{7/3}}\frac{a}{A^{1/3}} \frac{1}{r}\left( ds^2_{AdS_3} + 4~ ds^2_{S_3^3} \right) +  \frac{1}{2^{4/3}} A^{2/3}  \frac{1}{r^2} dr^2 +  \frac{1}{2^{4/3}} A^{2/3} ds^2_{S^4},
\e
where
\be
ds^2_{S^4} = d\theta^2 + \sin^2(\theta) ds^2_{S_2^3}.
\e
The metric \eqref{eq:M5.metric} is very similar to an $AdS_7 \times S^4$, and becomes that if one replaces the $AdS_3 \times S^3$ in parentheses by $\mathbf R^6$. So it is natural to interpret the solution as M5-branes wrapped on $AdS_3 \times S^3$, as we verify below. One may wonder if this solution is really $AdS_7 \times S^4$ in disguise, which is not possible because of the no-go theorem of \cite{D'Hoker:2008wc} that prevents the existence of solutions with asymptotic $AdS_7 \times S^4$ regions preserving the supergroup $OSp(4|2,\mathbb{R})\otimes OSp(4|2,\mathbb{R})$. To see that indeed this is not the case we perform the change of coordinates $r \rightarrow \frac{a}{2A}e^{2r}$ to get
\be  \label{eq:M5.metric2}
ds^2 \rightarrow  R^2_{S^4} \left[ e^{2 r}\left( ds^2_{AdS_3} + 4~ ds^2_{S_3^3} \right) +  4 ~ dr^2 + ds^2_{S^4} \right] ,
\e
with $R^2_{S^4} =2^{-4/3} A^{2/3}$. This should be compared with the $AdS_7 \times S^4$ metric in the appropriate limit. We write the $AdS_7$ metric as an $AdS_3 \times S^3$ fibration over a ``radial'' coordinate $r$
\be
ds^2_{AdS^7 \times S^4}=  R^2_{S^4} \left[4  \cosh^2(r)ds^2_{AdS_3} + 4 \sinh^2(r)ds^2_{S^3} + 4 ~ dr^2+ ds^2_{S^4} \right],
\e
and take the limit $r \rightarrow \infty$ to get
\be
ds^2_{AdS_7 \times S^4} \rightarrow   R^2_{S^4} \left[e^{2 r}\left( ds^2_{AdS_3} + ds^2_{S^3} \right) +  4 ~ dr^2+ ds^2_{S^4} \right].
\e
This metric differs from the metric \eqref{eq:M5.metric2} by a relative factor of 4 in $ds^2_{S^3}$, showing that \eqref{eq:M5.metric2} is not an $AdS_7 \times S^4$ space.

To compute the 4-form field strength we need the vector $J_{u}$. It takes the form
\be
J_u \rightarrow 3 A e^{-i \theta} \sin^3(\theta)\frac{1}{r} + 3 \frac{1}{a}e^{-i \theta} \left[2 A^2 - 4 a C \cos(\theta)  - i a C \sin(\theta)\right]\sin^3(\theta).
\e
The components of the 4-form field strength become
\be
\begin{split}
g_{1u} \rightarrow  & - \dfrac{3}{2 \sqrt{2}}\frac{A^{1/2}}{a^{1/2}}\frac{1}{r^{1/2}}e^{-2i\theta} -  \frac{3}{2^4 \sqrt{2}} \frac{1}{a^{3/2} A^{1/2}} r^{1/2} e^{-i2\theta} \left[A^2 - (A^2 + a C) \cos(\theta) - i 2 A^2 \sin(2 \theta) \right],\\
g_{2u} \rightarrow &  - i  \dfrac{3}{2} \frac{1}{r}e^{-i\theta} \\
& -  i \frac{3}{2^3} \frac{1}{a A} e^{-i\theta} \left[3 A^2 + 12 a C \cos(\theta) + (2 A^2 -16 a C)  \cos(2 \theta) - i 4a C \sin(\theta) + i A^2 \sin(2\theta) \right],\\
g_{3u} \rightarrow & - \dfrac{3}{2 \sqrt{2}}\frac{A^{1/2}}{a^{1/2}} \frac{1}{r^{1/2}}e^{-2i\theta} + \frac{3}{2^3 \sqrt{2}} \frac{1}{a^{3/2}A^{1/2}} r^{1/2} e^{-i2\theta} \left[2 a C\cos(\theta) -  A^2 \cos^2(\theta) \right].\\
\end{split}
\e
%
The 4-form flux through $S^4 \simeq S^3_2 \times_f [0,\pi]_{\theta}$ is
\be
\int_{S^4} F_4 = 2\pi^2 A \equiv  (2\pi)^3 N_{5}.
\e
The form of the metric and the 4-form flux suggest the presence of a stack of $N_{5}$ M5-branes wrapping an $AdS_3 \times S_3^3$ subspace. A stack of $N_{5}$ M5-branes wrapping an $AdS_3 \times S_2^3$ subspace is obtained by considering the solution with $G \rightarrow \bar G$.

The solution carries non-trivial M2-brane charge for $C\neq 0$. The conserved 7-form flux through the non-trivial 7-cycle $\mathcal M_7 \simeq S_3^3 \times S^4$ is given by
\be
\int_{\mathcal M_7} \left( \ast F_4 + \frac{1}{2} C_3 \wedge F_4 \right)  = \frac{2\cdot 3^3}{ 5} \pi^4 a C \equiv (2\pi)^6 n_2.
\e
We interpret this result as the local solution describing M2-branes intersecting and/or ending on a stack of $N_5$ M5-branes, in the near-horizon limit of the M2-branes, and in the region close to the M5-branes. This local solution could be part of a global solution dual to M2-branes ending on M5-branes, in which $n_2$ M2-branes end on a specific stack of $N_5$ M5-branes. But it could also be part of a global solution dual to $N_2$ M2-branes intersecting and ending on M5-branes,
in which $n_2$ of the $N_2$ M2-branes end on a specific stack of $N_5$ M5-branes. If $C=0$, there is no M2-brane charge on the M5-branes and we interpret this result as M2-branes intersecting a stack of $N_5$ M5-branes.

\subsection*{Acknowledgements}

It is a pleasure to thank Ofer Aharony and Micha Berkooz for very illuminating discussions. This work was supported in part by an Israel Science Foundation center for excellence grant, by the German-Israeli Foundation (GIF) for Scientific Research and Development, and by the Minerva foundation with funding from the Federal German Ministry for Education and Research.


\end{document}